\documentclass[conference]{IEEEtran}
\IEEEoverridecommandlockouts
\usepackage{cite}
\usepackage{amsmath,amssymb,amsfonts}
\usepackage{algorithmic}
\usepackage{graphicx}
\usepackage{textcomp}
\usepackage{xcolor}
\usepackage{subcaption}
\usepackage{tikz}
\usepackage{pgfplots}
\usepgfplotslibrary{fillbetween,groupplots,colorbrewer}

\def\BibTeX{{\rm B\kern-.05em{\sc i\kern-.025em b}\kern-.08em
    T\kern-.1667em\lower.7ex\hbox{E}\kern-.125emX}}
\begin{document}

\title{Energy and Bandwidth Efficiency of Event-Based Communication\\
\thanks{This work was partly funded by the German Ministry of Education and Research (BMBF) under grant 16KIS1028 (MOMENTUM) and 16KISK003K (Open6GHub)}
}

\author{\IEEEauthorblockN{Christopher Willuweit,
		Carsten Bockelmann, Armin Dekorsy}
	\IEEEauthorblockA{Dept. of Communications Engineering,\\ University of Bremen, Bremen, Germany \\
		Email: \{willuweit, bockelmann, dekorsy\} @ant.uni-bremen.de }}
\maketitle

\begin{abstract}
Wireless sensor nodes need a drastically reduced technical complexity to fit constraints of future applications. Reducing complexity often results in a degradation of energy and bandwidth efficiency. An interesting new approach that promises to reduce both technical complexity and energy consumption is event-based communication (EBC). While practical low-complexity implementations of such systems have already been proposed, the general question of energy and bandwidth efficiency remains open. In this paper, we compare these between EBC and a system relying on classical uniform sampling. We show that EBC is indeed much more energy efficient, and this comes at the cost of bandwidth efficiency. Therefore EBC is particularly suitable in combination with ultra-wideband communication.
\end{abstract}

\begin{IEEEkeywords}
Low Power Wireless Sensors, Low Complexity Wireless Sensors, Event-Based Sampling, Event-Based Communication, Energy Efficiency, Bandwidth Efficiency
\end{IEEEkeywords}

\section{Introduction}
Wireless Sensor Networks (WSN) can be found in many of todays scientific and commercial fields, such as Internet-of-Things, where signals have to be measured and processed at different points in space. The sensing entities - the \textit{sensor nodes} - communicate in a wireless fashion, as this avoids the need for cabling. Characteristic requirements for the sensor node are small dimensions, cheap production and energy efficiency \cite{gun09,f16}. The first two points can be achieved by reducing the technical complexity, since the required chip area influences both the size and the price of the sensor node. However, this reduction often comes at the cost of energy and bandwidth efficiency. For example, analog transmission can already be accomplished with a single transistor, but the desire for energy and bandwidth efficiency, among other things, replaced such systems with digital ones a long time ago.

At the interface between the analog measurement signal and the digital domain, state of the art sensor nodes are based on the Whittaker-Shannon-Kotelnikov (WSK) sampling theorem. When the measurement signals have more structure than just band limitation, a signal specific digital source coding is commonly used to compress the data to be transmitted. But digital signal processing in general has a very high complexity, i.e. System-on-Chips for wireless sensing often contain millions of transistors \cite{hnl08}. There are currently many research areas focussing on capturing the signal structure directly at sampling, such as analog-to-information conversion or compressed sensing \cite{ver15}.
Another interesting approach of capturing a signals structure below the WSK rate is \textit{event-based sampling} (EBS) \cite{rze14,mis16}. For structured signals it can produce sampling rates way below the WSK rate. First approaches for combining the EBS framework with communication to get an \textit{event-based communication} (EBC) have been made \cite{kli15,kli16}. Besides the sampling, also the communication hardware is significantly reduced in complexity when compared to state-of-the-art methods, e.g. multi-carrier transmission. But one could surely imagine a WSK based sensor node that is reduced to minimum technical complexity. The question is, if EBC is still not only less complex but also more energy and bandwidth efficient than this reduced WSK node. This question will be answered in this paper.

\begin{figure}[h]
	\centering
	\includegraphics[width=0.8\linewidth]{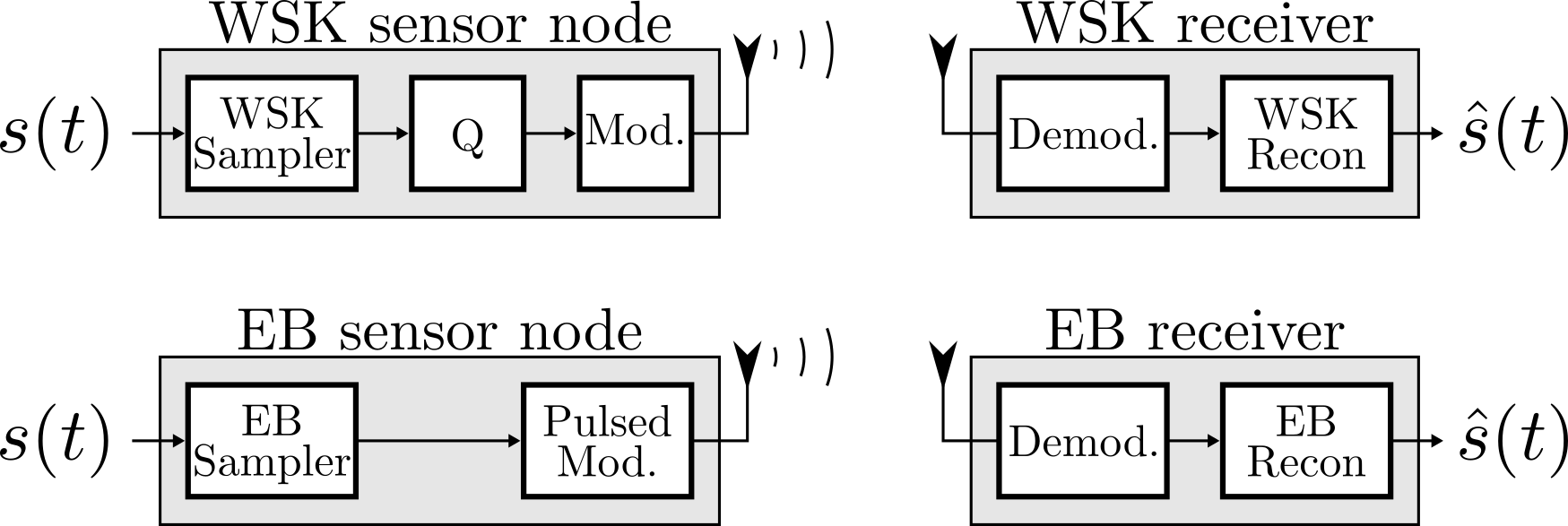}
	\caption{The two low complexity systems for wireless sensing we are comparing. Top: A  digital communication utilizing WSK sampling, quantization (Q) and digital modulation. Bottom: The proposed event-based communication (EBC) using an EB sampler generating nonuniform samples which are then directly modulated onto transmission pulses.}
	\label{fig:intro}
\end{figure}

In Fig.\,\ref{fig:intro} the structures of the two approaches are shown. The WSK setup consists of a uniform sampler and a quantization. The resulting bits are then modulated onto symbols. The EBC setup consists of an event-based sampler, generating nonuniform samples, which are then modulated onto symbols. Due to the nonuniform timing of these symbols we call them \textit{pulses}. In both setups the respective receiver estimates the measurement signal. 

We will first analyze which of the many possible event-based samplers are suitable and it's implications on the communication signals. Then we will parameterize the comparative WSK setup in a fair way and subsequently compare the transmit energy  and bandwidth demands of both setups.

\subsection{Related work}
The signal class we use in this work to descibe the \textit{structure} of measurement signals was originally proposed by\cite{cpl85} and further analyzed in \cite{wei06}. Research about the combination of this signal class with event-based sampling was driven by the control community. \cite{mis16} provides a good overview and concrete techniques for signal sampling, processing and reconstruction.

The event-based sampling we use, \textit{send-on-delta} (SOD) sampling, was used for wireless sensing in \cite{kli15}. The authors proposed EBC based on ultra wideband (UWB) frequency-shift keying. However they have focussed primarily on practical implementation of the communication, not so much on the influence of signal structure on performance. The system was later extended by an event-based error correction in \cite{kli16} and \cite{kli16b}.
\subsection{Contributions}
After introducing a concept of \textit{varying-bandwidth} signals and event-based sampling we show that the SOD sampler has a minimum time distance between two samples. We prove that this is true not only for classically band limited signals, but also for varying-bandwidth signals and derive concrete statements about the transmission bandwidth of EBC from it. 

To show the advantages of this system over classical approaches, we propose a low complexity WSK based sensor node as a comparison.

Event-based approaches are particularly suitable for signals with high variations in bandwidth. We therefore present a test signal class to generate random signals with different bandwidth variations.

While perfect reconstruction is well known for WSK sampling, this is not the case for event-based methods. Therefore, we present a simple and fast reconstruction method. Reconstruction accuracy and required transmit energy and bandwidth are compared for the WSK and EBC approach using numerical simulations. 

We elaborate the general connection between bandwidth variation and necessary number of samples and thus energy consumption. We show that event-based communication has a clear advantage over WSK systems in terms of energy efficiency at the expense of bandwidth efficiency.
\section{Event-based sampling}
The idea of classic WSK sampling is to sample a signal $s(t)$ at equidistant time points. The sampling times are only signal dependent in that they are based on the maximum frequency component of the signal. In contrast, the sampling times for \textit{event-based sampling} are directly dependent on the signal. Those samples can then be interpreted as events (e.g. \textit{the signal crossed a certain amplitude level}), hence the name. There is a variety of different sampling methods. Most of them require only a minimal hardware effort.
\subsection{Signal model}
\label{ssec:sigmal_model}
A signal framework that provides indications of both optimal sampling and reconstruction is \textit{time-warping}, which was proposed by Clark, Palmer, and Lawrence \cite{cpl85} already in 1985. It is an extension of the classic WSK sampling theorem. We start with a signal $\tilde{s}(\tau)$, which is band limited:
\begin{align}
\forall |f| > \tilde{B}: \quad \int_{-\infty}^{\infty} \tilde{s}(\tau) e^{-j2\pi f \tau} d\tau = 0   .
\label{eqn:s_tilde_BL}
\end{align}
We define a monotonically increasing \textit{warping function} $\gamma(t)$ and define the signal $s(t)$ as
\begin{align}
 s(t) = \tilde{s}(\gamma(t)).
\end{align}
Without loss of generality we assume that $\tilde{B}=1$. From the WSK sampling theorem we know, that we can perfectly reconstruct $\tilde{s}(\tau)$ from its uniform samples at $\tau_n = \frac{n}{2}, n\in\mathbb{N}$. Thus, when $\gamma(t)$ is known, we can also reconstruct $s(t)$ from its samples at $t_n = \gamma^{-1} (\tau_n)$. Except for the special case of a linear $\gamma(t)$, these samples are not uniformly distributed in $t$, but are concentrated in regions where $\frac{\delta \gamma(t)}{\delta t}$ is large and scarce where $\frac{\delta \gamma(t)}{\delta t}$ is small. Thus the \textit{local sampling rate} is proportional with the derivative of $\gamma (t)$. As signal bandwidth and sampling rate are directly proportional in classical WSK sampling, this leads us to a definition for an \textit{instantaneous} bandwidth
\begin{align}
W(t) := \frac{1}{2} \frac{\delta \gamma (t)}{\delta t},
\end{align}
which is purely positive. We call a signal $s(t)$ defined in this way a \textit{varying bandwidth signal}.

Note that the definition of $W(t)$ is very different from a signal bandwidth defined by the Fourier integral (as in eqn.(\ref{eqn:s_tilde_BL})). A varying-bandwidth signal has an infinite bandwidth $B$ unless $W(t)$ is a constant function. Nevertheless, a varying-bandwidth signal with maximum instantaneous bandwidth $W_\text{max} := \max_{t\in \mathbb{R}} W(t)$ can be approximated with a classically band limited signal with bandwidth $B = W_\text{max}$.

Time-warping suggests that many signals might be represented with less samples, than the WSK-theorem suggests. The number of samples per second to describe a signal $s$ is given not by $W_\text{max}$ but by the mean instantaneous bandwidth $\overline{W(t)}$. As soon as $\overline{W(t)} < W_\text{max}$, the signal will have less degrees of freedom than samples generated by WSK sampling at $2\cdot W_\text{max}$, thus it has structure beyond band limitation. For further reading on this concept we refer to \cite{cpl85} and \cite{mis16}.

Furthermore we assume that the measurement signal amplitudes are bounded
\begin{align}
|s(t)| < s_\text{max},
\end{align}
which is a  natural feature of real-world signals. We need the bound $s_\text{max}$ to calculate quantization errors of the WSK system and we will see in the next sections that it also gives us important information about the EBC systems transmission bandwidth.

\subsection{Send-on-Delta sampling}
If we want to sample and reconstruct varying-bandwidth signals we have two problems: First, the instantaneous bandwidth $W(t)$ must be known for the reconstruction. Second, a practical sampler that samples exactly at $\gamma^{-1}(\tau_n)$ is difficult to imagine. In \cite{rze14} it is shown that a sampler that generates an instantaneous sampling rate which is proportional to the instantaneous bandwidth is sufficient for signal reconstruction. The authors show this for so-called level-crossing samplers. These sample the signal $s(t)$ when it crosses several pre-defined amplitude levels. However, for reasons  explained in Chapter III, level-crossing samplers are not suitable for the low complexity event-based communication that we propose. Instead we use the highly related SOD sampler.

\begin{figure}
	\centering
	\begin{tikzpicture}
\begin{axis}[
name=a,
width=\linewidth, 
height=5cm,
xmin=0.25, xmax=0.35,
ymin=-2, ymax=3,
scaled x ticks=false,
xlabel={Time [s]},
scaled y ticks=false,
ylabel={Amplitude},
ylabel near ticks,
grid style={line width=.1pt, draw=gray!10},
major grid style={line width=.2pt,draw=gray!50},
grid=both,
legend pos = north west,
]
\addplot[color=blue,line width=1pt,no markers] 
table [x=t,y=s,col sep=comma] {pics/sod/signal.csv};
\addlegendentry{Signal}
\addplot[color=black,dashed,line width=1pt,no markers,forget plot] 
table [x=t,y=l1,col sep=comma] {pics/sod/levels.csv};
\addplot[color=black,dashed,line width=1pt,no markers,forget plot] 
table [x=t,y=l2,col sep=comma] {pics/sod/levels.csv};
\addplot[color=black,dashed,line width=1pt,no markers,forget plot] 
table [x=t,y=l3,col sep=comma] {pics/sod/levels.csv};
\addplot[color=black,dashed,line width=1pt,no markers,forget plot] 
table [x=t,y=l4,col sep=comma] {pics/sod/levels.csv};
\addplot[color=black,dashed,line width=1pt,no markers,forget plot] 
table [x=t,y=l5,col sep=comma] {pics/sod/levels.csv};
\addplot[color=black,dashed,line width=1pt,no markers,forget plot] 
table [x=t,y=l6,col sep=comma] {pics/sod/levels.csv};
\addplot[color=black,dashed,line width=1pt,no markers,forget plot] 
table [x=t,y=l7,col sep=comma] {pics/sod/levels.csv};
\addplot[color=black,dashed,line width=1pt,no markers] 
table [x=t,y=l8,col sep=comma] {pics/sod/levels.csv};
\addlegendentry{Levels}
\addplot[color=green!50!black,line width=1pt,mark=*,only marks] 
table [x=t,y=s,col sep=comma] {pics/sod/events.csv};
\addlegendentry{Samples}
\end{axis}
\draw [decorate,decoration={brace,amplitude=3pt},xshift=-0cm,yshift=0pt]
(0.0,1.65) -- (0.0,2.27)
node [black,midway,xshift=-0.4cm]
{\footnotesize $\Delta_L$};
\end{tikzpicture}	
	\caption{An example varying bandwidth signal $s(t)$ (blue graph). It's instantaneous bandwidth (highlighted in Fig.\ref{fig:testsignal_bws}) is monotonically rising in the time frame depicted here. Also shown are the SOD levels (dashed lines with $\Delta_L=\frac{8}{9}$) and the SOD samples (green) generated. It can be seen that the density of SOD samples is rising with the instantaneous bandwidth.}
	\label{fig:sod}
\end{figure}
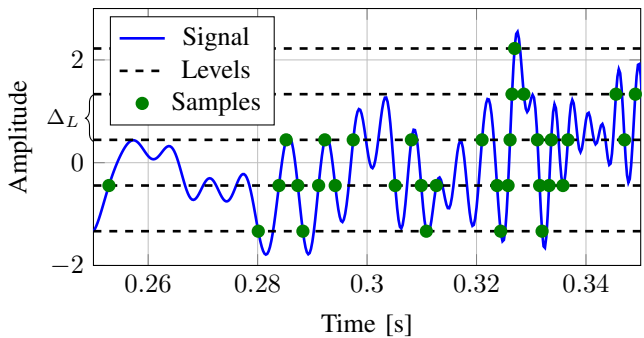

The concept of SOD sampling is used in many different areas from signal theory to computer science. Especially in communication it is used on different layers of abstraction, which is why the term SOD is highly ambiguous. To be clear, we will describe our definition of SOD: First, a set of equidistant amplitude levels with distance $\Delta_L$ is defined. The first level-crossing of a signal will generate a sample at the crossing. The next sample is triggered, when one of the two neighboring levels is crossed. Thus, samples are generated when an amplitude distance of $\Delta_L$ to the last sample is reached. This is visualized in Fig.\ref{fig:sod} with an examplary varying-bandwidth signal. 
When comparing the generated samples to the signals instantaneous bandwidth (shown in Fig.\ref{fig:testsignal_bws}), it can be seen that the density of samples rises with the instantaneous bandwidth. This is exactly the behaviour described in the last paragraph.

SOD samples can be coded in a very efficient way using just the information if the current sample is one level above or one level below the last sample, which leads to only one bit per sample. This is also where EBC gets its name: Signals are not represented by samples but by events (e.g. the signal has increased by $\Delta_L$). Obviously, if the signal at least once crosses all the levels, which we assume in this work, the actual samples can be derived uniquely from the event sequence.
\subsection{Events for bounded Signals}
The Bernstein/Zygmund theorem \cite{pinsky} shows, that band limited signals that are bounded, also have a bounded derivative. Applied to our signal model, this means that:
\begin{align}
\left| \frac{\delta \tilde{s}(\tau)}{\delta \tau} \right| \leq 2\pi \tilde{s}_\text{max}\tilde{B},
\end{align}
where $\tilde{B}$ is the bandwidth of $\tilde{s}$ with respect to $\tau$ and $\tilde{s}_\text{max}$ is its maximum amplitude. In the appendix we show a proof, that also the derivatives of bounded varying-bandwidth signals $s(t)$ are bounded:
\begin{align}
\left| \frac{\delta s(t)}{\delta t} \right| \leq 2\pi s_\text{max} W(t).
\end{align}
This is an interesting result, because as mentioned earlier, varing-bandwidth signals are not band limited in the classical sense. Nevertheless, Bersteins/Zygmunds inequality still holds. For the remainder of this work, the global bound over all $t$ is sufficient:
\begin{align}
\left| \frac{\delta s(t)}{\delta t} \right| \leq 2\pi s_\text{max} W_\text{max}.
\end{align}
We can directly apply this to get a lower bound $T_\text{LB}$ on the time between two SOD events, since the signal has a bounded rate to travel the needed amplitude distance $\Delta_L$ and trigger two consecutive samples:
\begin{align}
T_\text{LB} \geq \frac{\Delta_L}{2\pi s_\text{max} W_\text{max}}.
\label{eqn:T_min_bound}
\end{align}

\section{Event-based communication}
As stated in the introduction, our goal is to transmit the events by the use of transmit pulses directly after they occur. We do not focus on the specifics of these pulses but use some general thoughts about pulsed transmission in combination with SOD, which we will address in this chapter:
\begin{itemize}
	\item Each pulse has to have a certain energy to be detectable. 
	\item Consecutive pulses must have a minimum distance to be distinguishable.
\end{itemize}
To be detectable a pulse has to have an energy that is above the noise floor. We assume that the pulses of the EBC system (each representing a SOD sample encoded in a single bit) require the same amount of energy as the symbols of the WSK system to be detectable. This is of course only valid if the WSK system also just transmits a single bit per symbol. Therefore we assume that WSK must transmit $N_\text{bits}$ symbols per sample resulting in a symbol rate of $r_\text{symbol} = N_\text{bits}\cdot f_s$. Thus we establish comparability without having to analyze both systems in baseband. It allows us to compare directly from the number of samples generated per time, in both the WSK and EBC systems, the transmit powers required without knowing the concrete necessary symbol or pulse energy required. Therefore we define the relative transmit power as the ratio of event rate of the EBC system and symbol rate of the WSK system:
\begin{align}
p_\text{rel} = \frac{r_\text{event}}{r_\text{symbol}} = \frac{r_\text{event}}{N_\text{bits}\cdot f_s} .
\end{align}

For the relative transmission bandwidth of the EBC system we analyze the durations of WSK symbols and EBC pulses. The duration of a WSK symbol is given by $\frac{1}{N_\text{bits}f_s}$. For the EBC system we have a varying pulse rate. Nevertheless we assume a fixed pulse duration to keep EBCs low hardware complexity. Thus we have to design the pulse duration such that it does not exceed the minimum time distance $T_\text{min}$ between two events in a given measurement signal to prevent interference between consecutive pulses. This distance is lower-bounded by $T_\text{LB}$, as shown in the last chapter. Since the duration of a pulse is inversely proportional to it's bandwidth, we can formulate the relative transmission bandwidth of EBC, again without having to specify concrete pulse shapes:
\begin{align}
b_\text{rel} = \frac{1}{T_\text{min}\cdot N_\text{bits}\cdot f_s},
\end{align}
In the worst case, $T_\text{min} = T_\text{LB}$ the relative transmission bandwidth cannot be smaller than
\begin{align}
b_\text{rel,worst} = \frac{1}{T_\text{LB} \cdot N_\text{bits}\cdot f_s} = \frac{2\pi s_\text{max} W_\text{max}}{\Delta_L N_\text{bits} f_s}.
\end{align}
For further reading on the detectability of unsynchronized, partly overlapping transmit pulses we refer to \cite{wil22}.

\subsection{Reconstruction}
The reconstruction algorithm for the WSK sampling is commonly known, see e.g. \cite{zay93}. A number of different reconstruction methods exist for event based sampling, many of which are tailored to specific signal classes, e.g. for electrocardiogram (ECG) signals \cite{rav14}. A more general approach for  varying-bandwidth signals can be found in \cite{rze14}. Since these methods are usually of high complexity, we use a different, more simple approach that does not depend on the signal class at all: We simply linearly interpolate between the SOD samples, that are derived from the SOD events. Therefore, all performances of the EBC system shown in this paper should be understood as worst-case performances that can most likely be outperformed by more sophisticated reconstruction methods. On the other hand, we will always choose the parameters of the WSK system to achieve the best performance.

\section{Numerical simulation}
\subsection{Test signals}
\label{ssec:test_signals}
\begin{figure}
	\begin{tikzpicture}
\begin{axis}[
name=a,
width=\linewidth, 
height=4.5cm,
xmin=0, xmax=1,
ymin=0, ymax=1200,
scaled x ticks=false,
xlabel={Time $t$ [s]},
scaled y ticks=false,
ylabel={$W(t)$ [Hz]},
ylabel near ticks,
grid style={line width=.1pt, draw=gray!10},
major grid style={line width=.2pt,draw=gray!50},
grid=both,
legend pos = north east,
cycle list/Dark2,
]

\addplot +[line width=1pt] 
table [x=t,y=w5,col sep=comma] {pics/example_bws/example_bws.csv};
\node (n1) at (85,100) {\footnotesize $\overline{W} = 925$ Hz};
\addplot +[line width=1pt] 
table [x=t,y=w4,col sep=comma] {pics/example_bws/example_bws.csv};
\node (n1) at (85,80) {\footnotesize $\overline{W} = 775$ Hz};
\addplot +[line width=1pt] 
table [x=t,y=w3,col sep=comma] {pics/example_bws/example_bws.csv};
\node (n1) at (85,60) {\footnotesize $\overline{W} = 625$ Hz};
\addplot +[line width=1pt] 
table [x=t,y=w2,col sep=comma] {pics/example_bws/example_bws.csv};
\node (n1) at (85,40) {\footnotesize $\overline{W} = 475$ Hz};
\addplot +[line width=1pt,forget plot] 
table [x=t,y=w1,col sep=comma] {pics/example_bws/example_bws.csv};
\node (n1) at (85,20) {\footnotesize $\overline{W} = 325$ Hz};

\addplot +[line width=3pt,forget plot] 
table [x=t,y=w,col sep=comma] {pics/example_bws/example_bws_highlighted.csv};

\end{axis}
\end{tikzpicture}
	\caption{Instantaneous bandwidths with five different mean instantaneous bandwidths $\overline{W}$ used to generate our test signals. The highlighted region of the lower graph corresponds to the instantaneous bandwidth of the examplary signal in Fig.\ref{fig:sod}.}
	\label{fig:testsignal_bws}
\end{figure}
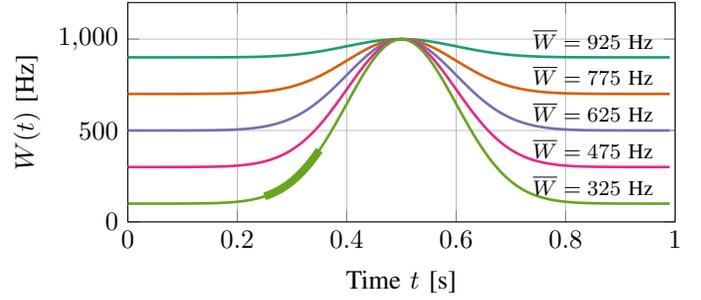

%
For comparing EBC and WSK performances we need to generate variable bandwidth test signals. To easily convert between energy and power, we choose a signal length of one second. First we determine the instantaneous bandwidths $W(t)$ of the test signals. Since we want to vary the structuredness of our signals, they should have the mean instantaneous bandwidth $\overline{W}$ as a parameter. Additionally they should vary smoothly, be purely positive and have a maximum $W_\text{max}$ that is constant w.r.t. $\overline{W}$. Therefore we choose an offset Gaussian function with $W_\text{max}=1$\,kHz. The standard deviation of the Gaussian is arbitrarily chosen to one tenth of the signal duration or 100\,ms:
\begin{align}
W(t) = \frac{4}{3} (\overline{W} - 250) + \left( \frac{4000}{3}-\frac{4}{3} \overline{W} \right) \cdot e^{-\frac{(t-0.5)^2}{0.02}}.
\end{align}
With these parameters, the mean instantaneous bandwidth is restricted to
\begin{align}
250\,\text{Hz} < \overline{W} < 1000\,\text{Hz}.
\end{align}
In this work we analyze five different $\overline{W}$: [325,475,625,775,925]\,Hz. The corresponding $W(t)$ are shown in Fig.\ref{fig:testsignal_bws}.

We now have to assign a number of amplitudes (see section \ref{ssec:sigmal_model})  that is given by $2 \overline{W}$. We choose them from a standard normal distribution $\mathcal{N}(0,1)$, resulting in a signal $s(t)$ with a mean power of one. To ensure boundedness, we only consider realizations for which $|s(t)| \leq s_\text{max} = 4$.

\subsection{Parameterization}
The WSK system has two free parameters: the sampling frequency $f_s$ and the number $N_\text{bits}$ of bits per sample for the quantization. Prior to WSK sampling an anti-aliasing lowpass\footnote{Butterworth design, order 8} with cutoff frequency of $\frac{f_s}{2}$ is applied to $s(t)$. The sampling frequencies were set to a range below and above $2\cdot W_\text{max} = 2\,$kHz to cover both undersampling and oversampling:
\begin{align}
	f_s = 2 \cdot W_\text{max} \cdot N_\text{os}, \; N_\text{os} \in \left[ 0.2,0.3,...,1.9,2 \right].
\end{align}
The quantization parameter $N_\text{bits}$ was chosen to be
\begin{align}
N_\text{bits} \in \left[ 3,4,...,8 \right].
\end{align}

The EBC system has a single parameter, the level distance $\Delta_L$. We figured the following ranges of the parameter to yield similar rates to those of the WSK system:
\begin{align}
\Delta_L = \frac{2 s_\text{max}}{(N_L-1)} = \frac{8}{N_L-1}, \; N_L \in \left[ 10,15,...,95,100 \right],
\end{align}
where $N_L$ describes the number of levels in the signal amplitude range ($-4$ to $4$).

\subsection{Evaluation}
All the results are based on means of $M=100$ randomly chosen test signals $s_m(t), m \in [1,...,M]$  for each of the five different instantaneous bandwidths $W(t)$. 
To get the final results on energy and bandwidth efficiency, we sample and reconstruct each test signal with the $114$ different combinations of $N_\text{bits}$ and $f_s$ parameters for WSK and $20$ different $\Delta_L$ for EBC.

The symbol rates for the WSK system can be calculated analytically by $r_\text{symbol} = N_\text{bits} \cdot f_s$. The event rates for the EBC system are evaluated numerically. Since the observed time interval is one second, we simply calculate the mean number of occured events for all 100 realizations. The same holds for the minimum distance between two events, which will be used to calculate the transmission bandwidth.

To calculate errors we use the normalized mean squared error (NMSE) of reconstructed signals $\hat{s}_m(t)$
\begin{align}
\text{NMSE} = \frac{1}{N \cdot M}\sum_{m=1}^{M} \sum_{n=0}^{N} \frac{\left|s_m(n/N) - \hat{s}_m(n/N)\right|^2}{\left|s_m(n/N)\right|^2}
\end{align}
for the $M = 100$ signal realizations and $N = 16000$ time points, which corresponds to an oversampling factor of 8 with respect to $W_\text{max}$.

\begin{figure}[h]
	\begin{tikzpicture}
\begin{axis}[
name=a,
width=\linewidth, 
height=4.5cm,
ymode = log,
xmin=0, xmax=25000,
ymin=0.0001, ymax=0.1,
scaled x ticks=false,
xlabel={Symbol or event rate [$\frac{1}{\text{s}}$]},
scaled y ticks=false,
ylabel={NMSE},
ylabel near ticks,
grid style={line width=.1pt, draw=gray!10},
major grid style={line width=.2pt,draw=gray!50},
grid=both,
legend pos = south west,
]
\addplot[color=black,dotted,line width=1pt,forget plot] 
table [x=b3,y=q3,col sep=comma] {pics/example_transmission/nyq_data_2.csv};
\addplot[color=black,dotted,line width=1pt,forget plot] 
table [x=b4,y=q4,col sep=comma] {pics/example_transmission/nyq_data_2.csv};
\addplot[color=black,dotted,line width=1pt,forget plot] 
table [x=b5,y=q5,col sep=comma] {pics/example_transmission/nyq_data_2.csv};
\addplot[color=black,dotted,line width=1pt,forget plot] 
table [x=b6,y=q6,col sep=comma] {pics/example_transmission/nyq_data_2.csv};
\addplot[color=black,dotted,line width=1pt,forget plot] 
table [x=b7,y=q7,col sep=comma] {pics/example_transmission/nyq_data_2.csv};
\addplot[color=black,dotted,line width=1pt] 
table [x=b8,y=q8,col sep=comma] {pics/example_transmission/nyq_data_2.csv};
\addlegendentry{Nyquist}

\addplot[color=blue,line width=1pt] 
table [x=b,y=NMSE,col sep=comma] {pics/example_transmission/sod_data_2.csv};
\addlegendentry{EBC}

\node (n1) at (100,-1.5) {\footnotesize 3\,bit};
\node (n1) at (140,-3) {\footnotesize 4\,bit};
\node (n1) at (180,-4.5) {\footnotesize 5\,bit};
\node (n1) at (220,-5.9) {\footnotesize 6\,bit};
\node (n1) at (220,-7.25) {\footnotesize 7\,bit};
\node (n1) at (220,-8.6) {\footnotesize 8\,bit};
\end{axis}

\end{tikzpicture}
	\caption{WSK and EBC performance for $\overline{W} = 475\,$Hz. The different symbol and event rates are achieved by varying $\Delta_L$ and $f_s$.}
	\label{fig:example_performance}
\end{figure}
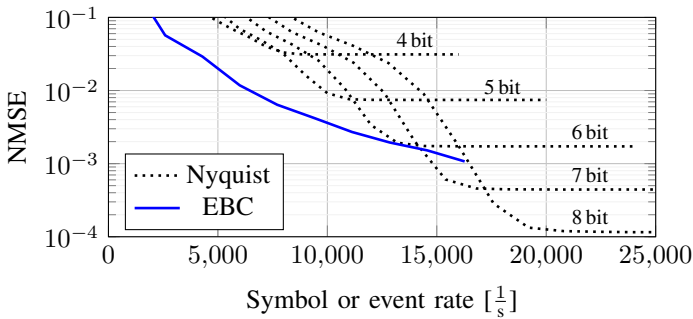

The perfomance of both systems for $\overline{W} = 475\,$Hz is shown in Fig.\ref{fig:example_performance}.
It can be seen that for this mean instantaneous bandwidth, EBC has an advantage over WSK in certain NMSE and rate regions. For the following analysis of transmission power and bandwidth we chose the WSK parameters $N_\text{bits}$ and $f_s$ that give the lowest symbol rate for a given NMSE and compare it to the EBC performance. This is done for each NMSE individually. For example we chose a 4\,bit quantizer for an NMSE of $10^{-1}$ and a 7\,bit quantizer for an NMSE of $10^{-3}$.
\section{Results}
\subsection{Energy efficiency}
\begin{figure}[h]
	\begin{tikzpicture}
\begin{axis}[
name=a,
width=0.8\linewidth, 
height=5cm,
xmode = log,
xmin=0.001, xmax=0.1,
ymin=0.25, ymax=1.75,
scaled x ticks=false,
xlabel={Target NMSE},
scaled y ticks=false,
ylabel={Rel. Tx-Power $p_\text{rel}$},
ylabel near ticks,
grid style={line width=.1pt, draw=gray!10},
major grid style={line width=.2pt,draw=gray!50},
grid=both,
legend pos = outer north east,
cycle list/Dark2,
]
\addlegendimage{empty legend}
\addlegendentry{\hspace{-0.5cm}$\overline{W}$ [Hz]:}
\addplot +[line width=1pt] 
table [x=t,y=E,col sep=comma] {pics/plots/plot_data/tx_energy_4.csv};
\addlegendentry{$925$}
\addplot +[line width=1pt, dashed] 
table [x=t,y=E,col sep=comma] {pics/plots/plot_data/tx_energy_3.csv};
\addlegendentry{$775$}
\addplot +[line width=1pt, dotted] 
table [x=t,y=E,col sep=comma] {pics/plots/plot_data/tx_energy_2.csv};
\addlegendentry{$625$}
\addplot +[line width=1pt, dashdotted] 
table [x=t,y=E,col sep=comma] {pics/plots/plot_data/tx_energy_1.csv};
\addlegendentry{$475$}
\addplot +[line width=1pt, loosely dashed] 
table [x=t,y=E,col sep=comma] {pics/plots/plot_data/tx_energy_0.csv};
\addlegendentry{$325$}
\end{axis}
\end{tikzpicture}
	\caption{Relative transmit power of the EBC system in dependence of the mean instantaneous bandwidth $\overline{W}$.}
	\label{fig:energy_eff}
\end{figure}
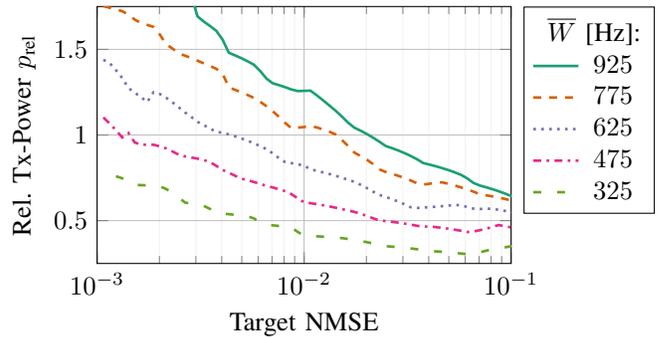

In Fig.\ref{fig:energy_eff} the relative transmit power for the EBC system is shown.
It can be seen that for the lowest $\overline{W}$, EBC needs less than half of the transmission power of WSK for NMSEs between $6 \cdot 10^{-3}$ and $3 \cdot 10^{-1}$. The advantage of EBC degrades with increasing mean instantaneous bandwidth $\overline{W}$. Very low NMSEs cannot be reached with EBC without drastically increasing the power consumption. This is of course also due to the simple reconstruction algorithm we used in this paper.
\subsection{Bandwidth efficiency}
\begin{figure}[h]
	\begin{tikzpicture}
\begin{axis}[
name=a,
width=0.8\linewidth, 
height=5cm,
xmode = log,
xmin=0.001, xmax=0.1,
ymin=4, ymax=10,
scaled x ticks=false,
xlabel={Target NMSE},
scaled y ticks=false,
ylabel={Rel. Tx-Bandwidth $b_\text{rel}$},
ylabel near ticks,
grid style={line width=.1pt, draw=gray!10},
major grid style={line width=.2pt,draw=gray!50},
grid=both,
legend pos = outer north east,
cycle list/Dark2,
]
\addlegendimage{empty legend}
\addlegendentry{\hspace{-0.5cm}$\overline{W}$ [Hz]:}
\addplot +[line width=1pt] 
table [x=t,y=B,col sep=comma] {pics/plots/plot_data/tx_bw_4.csv};
\addlegendentry{$925$}
\addplot +[line width=1pt, dashed] 
table [x=t,y=B,col sep=comma] {pics/plots/plot_data/tx_bw_3.csv};
\addlegendentry{$775$}
\addplot +[line width=1pt, dotted] 
table [x=t,y=B,col sep=comma] {pics/plots/plot_data/tx_bw_2.csv};
\addlegendentry{$625$}
\end{axis}
\end{tikzpicture}
	\caption{Relative transmit bandwidth of the EBC system in dependence of the mean instantaneous bandwidth $\overline{W}$. The performances for $\overline{W}=475$ and $\overline{W}=325$ are not shown because they are very close to that of $\overline{W}=625$}
\label{fig:bw_eff}
\end{figure}
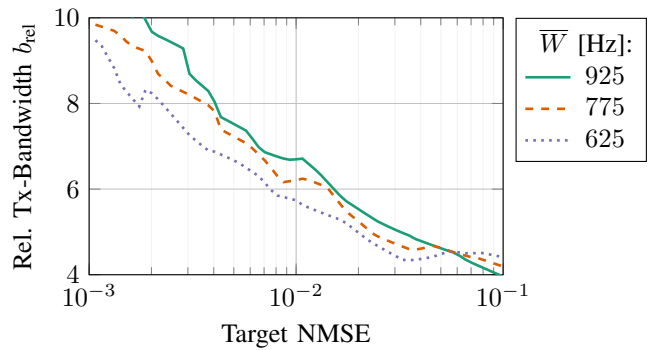

In Fig. \ref{fig:bw_eff} the relative transmission bandwidths $b_\text{rel}$ are depicted. These are based on the mean minimum distance $T_\text{min}$ between two SOD events of all test signals. The worst-case bandwidths $b_\text{rel,worst}$ are not shown for reasons of clarity, but they are about a factor of two above $b_\text{rel}$. 
The performance gets slightly better for more structured signals (smaller mean instantaneous bandwidth $\overline{W}$). Nevertheless the relative transmission bandwidth is clearly higher than one, which makes EBC less bandwidth efficient than a WSK approach. We find it unlikely that an EBC system will achieve the bandwidth efficiency of a WSK system even with much better reconstruction algorithms.
\section{Conclusion}
With EBC, we have presented a low complexity alternative to the currently very present WSK based wireless sensing. In addition to the complexity of sensor nodes, however, their energy and bandwidth efficiency also plays a major role, which we both examined.

We have shown that the EBC system can achieve much better energy efficiencies than the WSK system, especially when the measurement signals are highly structured. This is already evident with the very simple reconstruction algorithm we use. We expect that more sophisticated reconstruction algorithms will be able to improve this performance significantly. The improved energy efficiency comes at the expense of bandwidth efficiency, which makes the system particularly interesting for ultra-wideband communication, where bandwidth efficiency plays a minor role.

\appendix
\section*{Derivatives of VBW functions are bounded}
Assume a function $\mathbb{R} \to \mathbb{C}: \; \tilde{s}(\tau)$ with $\tau \in \mathbb{R}$ and that $\tilde{s}(\tau)$ is band limited to $\tilde{B}$:
\begin{align}
\forall |f| > \tilde{B}: \quad \int_{-\infty}^{\infty} \tilde{s}(\tau) e^{-j2 \pi f \tau} d\tau = 0   
\end{align}
and bounded in amplitude by $\tilde{s}_\text{max}$:
\begin{align}
\forall \tau \in \mathbb{R}: \; |\tilde{s}(\tau)| < \tilde{s}_\text{max}.
\end{align}
from the Bernstein/Zygmund theorem \cite{pinsky} follows, that the derivative of $\tilde{s}(\tau)$ is bounded:
\begin{align}
\left| \frac{\delta \tilde{s}(\tau)}{\delta \tau} \right| \leq 2\pi \tilde{s}_\text{max}\tilde{B}.
\end{align}
Let $\mathbb{R} \to \mathbb{R}: \;W(t)$ be a strictly positive and bounded function:
\begin{align}
\forall t \in \mathbb{R}: \; 0 < W(t) < W_\text{max}.
\label{eqn:W_is_positive}
\end{align}
Its antiderivative $\gamma(t)$ is thus a strictly monotonically increasing, differentiable function:
\begin{align}
\gamma(t) = \int_{-\infty}^{t} W(t^\prime) dt^\prime \; \Leftrightarrow \; \frac{\delta \gamma(t)}{\delta t} = W(t).
\label{eqn:W_vs_gamma}
\end{align}
Assume a function $s(t)$, that is the time-warped (with $\gamma(t)$) version of $\tilde{s}(\tau)$:
\begin{align}
s(t) = \tilde{s}(\gamma(t)).
\end{align}
Note that $\tilde{s}(\gamma(t))$ is only band limited w.r.t. $\gamma$, not necessarily w.r.t. $t$. 
Since time-warping does not change amplitude, also $s(t)$ is bounded:
\begin{align}
\forall t \in \mathbb{R}: \; |s(t)| < s_\text{max} = \tilde{s}_\text{max}.
\label{eqn:s_tilde_bounded}
\end{align}
We use the chain rule of differentiation:
\begin{align}
\frac{\delta s(t)}{\delta t} = \frac{\delta \gamma(t)}{\delta t} \cdot \frac{\delta \tilde{s}(\gamma)}{\delta \gamma}.
\end{align}
We know that $\frac{\delta \gamma(t)}{\delta t} > 0$ since it is monotonically increasing. Thus we can formulate the absolute value of the derivative as:
\begin{align}
\left| \frac{\delta s(t)}{\delta t} \right| = \left| \frac{\delta \gamma(t)}{\delta t}\right| \cdot \left| \frac{\delta \tilde{s}(\gamma)}{\delta \gamma} \right|.
\end{align}
The derivative of $\gamma(t)$ is equal to $W(t)$ (eqn. (\ref{eqn:W_vs_gamma})). Its absolute is also equal to $W(t)$, since it is strictly positive (eqn. \ref{eqn:W_is_positive}). Thus,
\begin{align}
\left| \frac{\delta s(t)}{\delta t} \right| = W(t) \cdot \left| \frac{\delta \tilde{s}(\gamma)}{\delta \gamma} \right|.
\end{align}
The right factor is bounded by the Bernstein/Zygmund bound, since $\tilde{s}(\gamma)$ is bounded and band limited w.r.t. $\gamma$:
\begin{align}
\left| \frac{\delta \tilde{s}(\gamma)}{\delta \gamma} \right| \leq 2\pi s_\text{max} \tilde{B}.
\end{align}
When we normalize $\tilde{B} = 1$,
\begin{align}
\left| \frac{\delta s(t)}{\delta t} \right| \leq 2\pi s_\text{max}\cdot W(t) \qquad \blacksquare
\end{align}

\end{document}